\documentclass[twocolumn,twoside,slac_two]{revtex4}
\usepackage{graphicx}
\usepackage{fancyhdr}
\begin{document}

\title{VHE $\gamma$-ray emission from the FSRQs observed by the MAGIC telescopes}
\author{E. Lindfors, K.Nilsson}\affiliation{Finnish Centre for Astronomy with ESO, University of Turku, Finland, contact: elilin@utu.fi} \author{U. Barres de Almeida, D. Mazin, D. Paneque, K. Saito} \affiliation{Max-Planck-Institut für Physik, Münich, Germany} \author{J. Becerra Gonzalez} \affiliation{Inst. de Astrofisica de Canarias, Spain and University of Hamburg, Germany} \author{K. Berger}\affiliation{Inst. de Astrofisica de Canarias, Spain} \author{G. De Caneva}\affiliation{DESY, Zeuthen, Germany} \author{C. Schultz} \affiliation{Universita di Padova and INFN, Italy} \author{J. Sitarek} \affiliation{IFAE, Edifici Cn., Campus UAB, Spain} \author{A. Stamerra}\affiliation{Universita di Pisa, and INFN Pisa, Italy} \author{F. Tavecchio} \affiliation{INAF National Institute for Astrophysics, Italy \\ on behalf of the MAGIC collaboration} \author{S.Buson}\affiliation{INFN and Universita di Padova, Italy} \author{F. D'Ammando}\affiliation{Univ. Perugia and INFN} \author{M. Hayashida} \affiliation{SLAC, California Institute of Technology, USA and Kyoto University, Japan\\ on behalf of the Fermi-LAT collaboration} \author{A. L\"ahteenm\"aki, M.Tornikoski}\affiliation{Aalto University Mets\"ahovi Radio Observatory} \author{T. Hovatta}\affiliation{Owens Valley Radio Observatory, California Institute of Technology, USA} 


%

%

\begin{abstract}
Among more than fifty blazars detected in very high energy (VHE, E$>$100GeV) $\gamma$-rays, only three belong to the subclass of Flat Spectrum Radio Quasars (FSRQs): PKS~1510-089, PKS~1222+216 and 3C~279. The detection of FSRQs in the VHE range is challenging, mainly because of their steep soft spectra in the GeV-TeV regime. MAGIC has observed and detected all FSRQs known to be VHE emitters up to now
and found that they exhibit very different behavior. The 2010 discovery of PKS 1222+216 (z = 0.432) with the fast variability observed, challenges simple one-zone emission models and more complicated scenarios have been proposed. 3C~279 is the most distant VHE $\gamma$-ray emitting AGN (z = 0.536), which was discovered by MAGIC in 2006 and detected again in 2007. In 2011 MAGIC observed 3C 279 two times: first during a monitoring campaign and later observations were triggered by a flare detected with {\it Fermi}-LAT. We present the MAGIC results and the multiwavelength behavior during this flaring epoch. Finally, we report the 2012 detection of PKS 1510-089 (z = 0.36), together with its simultaneous multiwavelength data from optical to $\gamma$-rays.
\end{abstract}

\maketitle

\thispagestyle{fancy}


\section{INTRODUCTION}

\begin{figure}
\centering
\includegraphics[width=85mm]{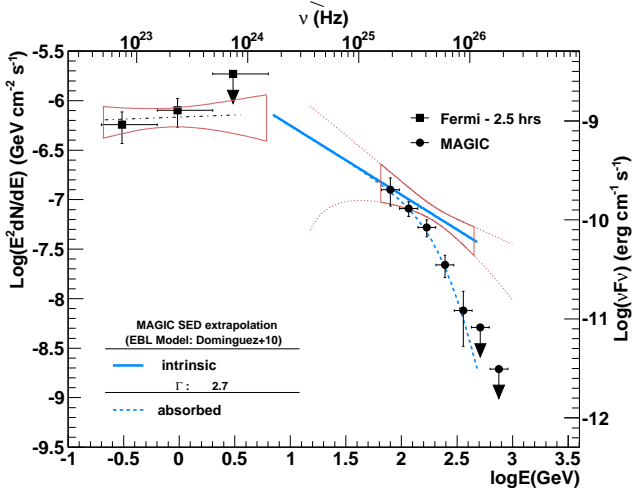}
\caption{PKS~1222+216: The high energy SED of the flare of 2010 June 17th. The retangular data points report the {\it Fermi}-LAT data of 2.5 hours observation centered around the 30 minutes observation by MAGIC (filled circles). The figure is adapted from \citep{1222}} \label{sed1222}
\end{figure}
\begin{figure}[t]
\centering
\includegraphics[width=85mm]{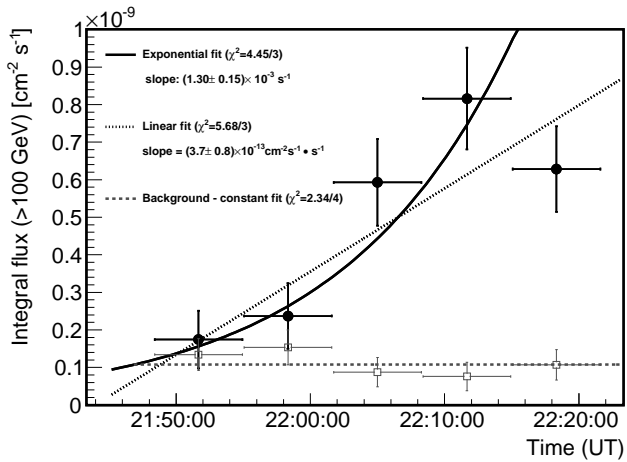}
\caption{PKS~1222+216: The lightcurve of the flare above 100\,GeV in 6 min bins. The figure adapted from \citep{1222}.} \label{lc1222}
\end{figure}

Blazars show variable emission in all wavebands from radio to
$\gamma$-rays, but in FSRQs the variability is typically larger in
amplitude. VHE $\gamma$-ray FSRQs are also more luminous in MeV-GeV
band than the VHE $\gamma$-ray emitting BL Lac objects and in general
also more distant. The spectral energy distribution (SED) of both subclass of blazars
show two peaks: first peak is generally attributed to synchrotron
emission and the second to inverse Compton scattering. Also
    hadronic mechanisms for producing the second peak has been
    proposed \citep[e.g.][]{bottcher}. In the case of FSRQs the first peak is typically in
the infrared regime while for BL Lacs it is between optical and hard
X-rays. The optical spectra of FSRQs show broad emission lines,
implying an existence of fastly moving gas clouds close (0.1 to 1
parsec) to the central engine, while BL Lacs show very weak or no
emission lines in their spectra. Because of these properties, FSRQs
were considered not to be good candidates to emit VHE $\gamma$-rays:
the low synchrotron peak frequency implies efficient cooling \citep[e.g.][]{ghisellini}, which
makes it difficult to produce VHE $\gamma$-ray emission, the broad
line region clouds absorb the $\gamma$-ray emission via pair
production and the large redshift also means stronger absorption of
VHE $\gamma$-rays by the extragalactic background light on their
travel to earth. However, three FSRQs have been detected by the
Imaging Air Cherenkov Telescopes (IACTs) and here we summarize the
recent results of FSRQ observations from the Major Atmospheric
Gamma-ray Imaging Cherenkov (MAGIC) Telescopes. MAGIC, operating in
stereo mode since Fall 2009, is two 17 m IACTs sited on the Canary
Island of La Palma at 2200 m a.s.l. Its energy threshold of 50\,GeV in
standard trigger mode is currently the lowest available, and it has a
sensitivity in stereo mode $<$ 0.8\% Crab Unit above 300 GeV in 50 hrs
of observations \citep{Performance}.

\section{PKS~1222+216}

The blazar PKS~1222+216 (4C +21.35) was discovered at VHE $\gamma$-rays by MAGIC on June 17th 2010 \citep{1222} during follow up observations triggered by high activity in high energy (HE) $\gamma$-rays in the {\it Fermi}-LAT range. MAGIC measured a soft spectrum with a photon index $\Gamma =3.75\pm 0.27_\mathrm{stat}$ $\pm$ 0.2$_\mathrm{sys}$  in the energy range from 70\,GeV to 400\,GeV. Complementing the measurements with {\it Fermi}-LAT observations, the simultaneous GeV and VHE (corrected for EBL absorption) spectra are consistent with a single power law with an index of $2.7\pm0.3$ between 3 GeV and 400 GeV, without a strong intrinsic cutoff (see Fig.~1). This strongly suggests that the $\gamma$-ray emission is located beyond the BLR. The VHE light curve (Fig.~2) shows a flux doubling time of the flare of the order of 10 minutes, among the fastest variations ever detected for sources of this class, implying a compact emission region, which is difficult to reconcile with a scenario where the $\gamma$-ray emission is not produced inside the BLR.

\begin{figure}[t]
\centering
\includegraphics[width=95mm]{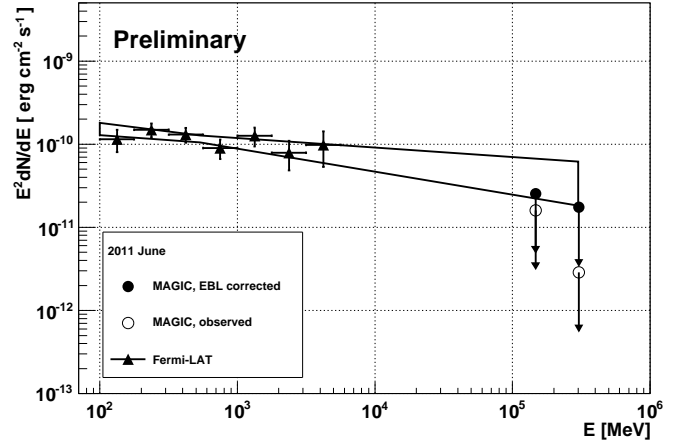}
\caption{Spectral energy distribution of 3C 279 measured with {\it Fermi}-LAT and MAGIC  in June 2011. The upperlimits from MAGIC are corrected for EBL absorption using \cite{dominguez}.} 
\label{sed3c279}
\end{figure}
\begin{figure}[t]
\centering
\includegraphics[width=85mm]{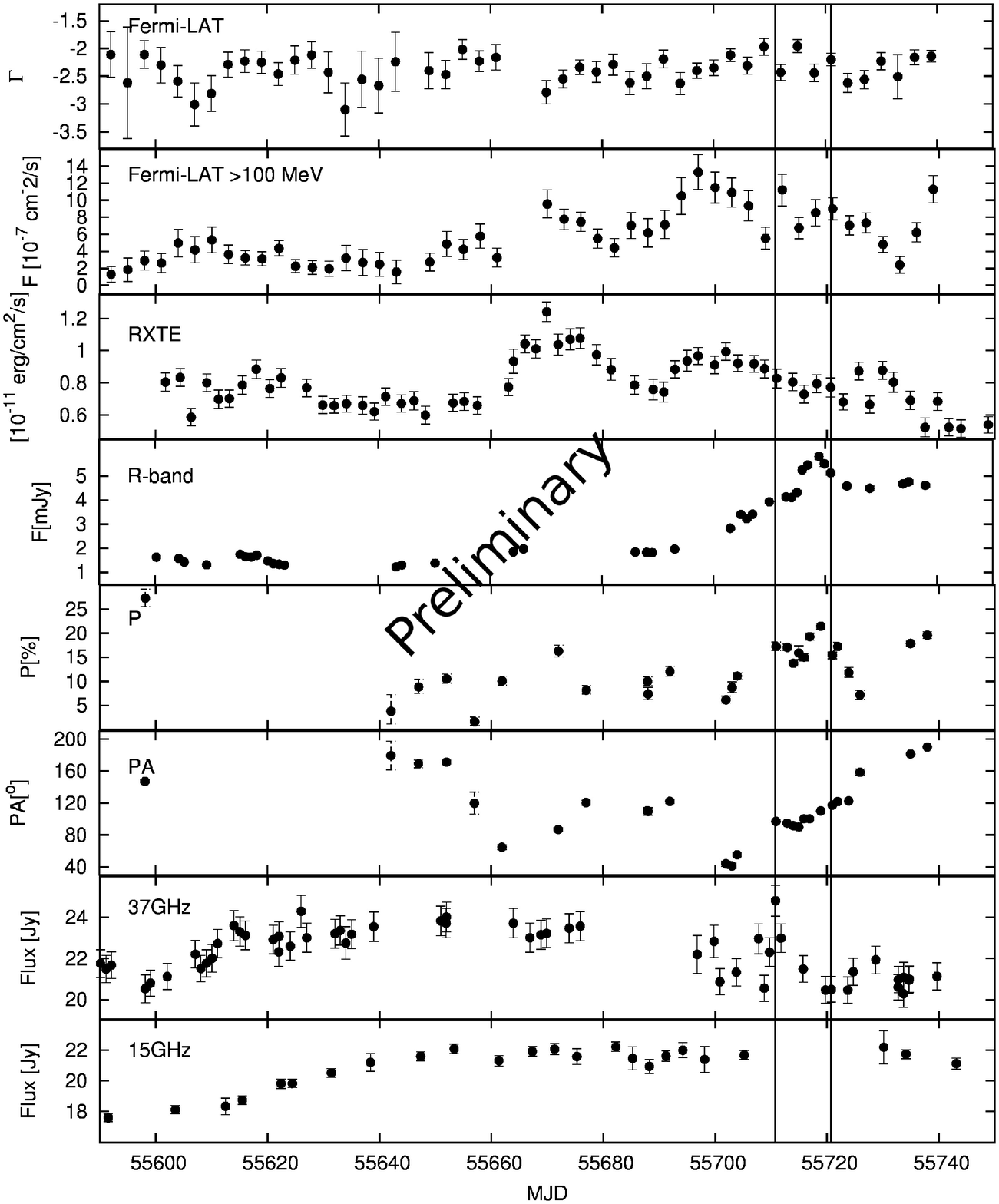}
\caption{The MWL light curve of 3C279 from radio to $\gamma$-rays. The vertical lines indicate the MAGIC observing epoch in June 2011.} \label{lc3c279}
\end{figure}

\section{3C~279}
3C~279 was discovered at VHE $\gamma$-rays by MAGIC on February 23rd 2006 \citep{Science} and redetected on January 17th 2007 \citep{3C279}. Both observations took place during optical high state, in particular 2007 observations during a major optical outburst (largest one observed in the past 10 years from the source). In the spring of 2011 3C~279 was showing increasing optical flux, as well as increased $\gamma$-ray flux as observed by {\it Fermi}-LAT, which triggered MAGIC observations of the source. The MAGIC observations did not yield in a detection, but provided tighter upper limits at VHE with respect to previous MAGIC observations performed in 2009 \citep{3C279}. These upper limits together with the simultaneous Fermi spectrum suggests a break in the spectra at $>10$ GeV.

The multiwavelength light curves from radio to $\gamma$-rays (Fig. 4) show complex behavior. In $\gamma$-rays there are three flares, the brightest one peaking on MJD 55698. In X-rays there are two flares which are simultaneous to $\gamma$-ray flares observed by {\it Fermi}-LAT. The optical outburst starts around the time when the second flare in $\gamma$- and X-rays is peaking and is accompanied by an increase of the optical polarization and a rotation of the polarization angle. While the MAGIC observations missed the largest $\gamma$-ray peak, there was still small $\gamma$-ray flare in this epoch whereas in X-rays the source was already decaying smoothly. The radio light curve indicates a fast flare at the beginning of the MAGIC observing epoch.

\section{PKS~1510-089}
Following the communication from the {\it Fermi}-LAT and AGILE collaborations, PKS~1510-089 was observed by MAGIC starting on February 3rd 2012. The observations continued until April 3rd 2012. These observations resulted in a detection of VHE $\gamma$-rays from the source with $>5 \sigma$ significance \citep{atel}. The source had been previously detected to emit VHE $\gamma$-rays by the H.E.S.S. telescope array in 2009 \citep{wagner} during a major optical outburst. The optical outburst observed in February-March 2012 was almost an order of magnitude smaller, but still the second brightest in the Fermi era. Unlike the other FSRQ detections by MAGIC, PKS~1510-089 signal comes from several nights. The analysis of the MAGIC data is ongoing and will be combined with radio (F-Gamma, Mets\"ahovi, OVRO, VLBA), optical (KVA, Liverpool), X-rays ({\it Swift}) and $\gamma$-rays ({\it Fermi}-LAT, AGILE) in the upcoming publication. Part of the multiwavelength light curves is shown in Fig. 5.

\begin{figure}[t]
\centering
\includegraphics[width=105mm, angle=-90]{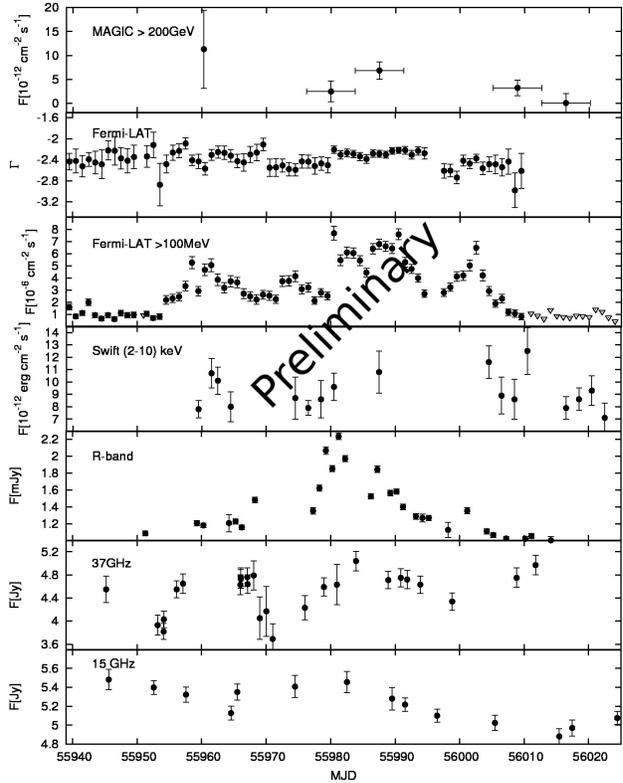}
\caption{The MWL light curve of PKS~1510-089 from radio to VHE $\gamma$-rays in spring 2012.} \label{lc1510}
\end{figure}

\section{SUMMARY}
The three known VHE gamma-ray emitting FSRQs, namely PKS~1222+216, 3C~279 and PKS~1510-089, have been observed and detected by MAGIC: 
\begin{enumerate}

\item The detections have all taken place during the activity in the lower energy 
regimes, but some of the observations in high optical and $\gamma$-ray states 
have only resulted in flux upper limits. 

\item The activity in the lower energy regimes has not followed the same patterns during the four VHE detections.  

\item The three sources show different time scales of variability in VHE gamma-rays.
\end{enumerate}

\section{Acknowledgments} \label{Ack}
We would like to thank the Instituto de Astrof\'{\i}sica de
Canarias for the excellent working conditions at the
Observatorio del Roque de los Muchachos in La Palma.
The support of the German BMBF and MPG, the Italian INFN, 
the Swiss National Fund SNF, and the Spanish MICINN is 
gratefully acknowledged. This work was also supported by the CPAN CSD2007-00042 and MultiDark
CSD2009-00064 projects of the Spanish Consolider-Ingenio 2010
programme, by grant DO02-353 of the Bulgarian NSF, by grant 127740 of 
the Academy of Finland,
by the DFG Cluster of Excellence ``Origin and Structure of the 
Universe'', by the DFG Collaborative Research Centers SFB823/C4 and SFB876/C3,
and by the Polish MNiSzW grant 745/N-HESS-MAGIC/2010/0.
The $Fermi$ LAT Collaboration acknowledges support from a number of
agencies and institutes for both development and the operation of the LAT
as well as scientific data analysis. These include NASA and DOE in the
United States, CEA/Irfu and IN2P3/CNRS in France, ASI and INFN in Italy,
MEXT, KEK, and JAXA in Japan, and the K.~A.~Wallenberg Foundation, the
Swedish Research Council and the National Space Board in Sweden.
Additional support from INAF in Italy and CNES in France for science
analysis during the operations phase is also gratefully acknowledged.
The Mets\"ahovi team acknowledges the support from the Academy of Finland
to our observing projects (numbers 212656, 210338, 121148, and others).
The OVRO 40-m monitoring program is
supported in part by NASA grants NNX08AW31G
and NNX11A043G, and NSF grants AST-0808050
and AST-1109911.


\end{document}